\documentclass[a4paper]{article}

\usepackage{INTERSPEECH2019}

\title{Deep Neural Network Embeddings with Gating Mechanisms for Text-Independent Speaker Verification}
\name{Lanhua You, Wu Guo, Lirong Dai, Jun Du}
\address{
National Engineering Laboratory for Speech and Language Information Processing\\
University of Science and Technology of China, Hefei, China
}
\email{lhyou@mail.ustc.edu.cn, \{guowu,lrdai,jundu\}@ustc.edu.cn}

\begin{document}

\maketitle
\begin{abstract}
In this paper, gating mechanisms are applied in deep neural network (DNN) training for x-vector-based text-independent speaker verification. First, a gated convolution neural network (GCNN) is employed for modeling the frame-level embedding layers. Compared with the time-delay DNN (TDNN), the GCNN can obtain more expressive frame-level representations through carefully designed memory cell and gating mechanisms. Moreover, we propose a novel gated-attention statistics pooling strategy in which the attention scores are shared with the output gate. The gated-attention statistics pooling combines both gating and attention mechanisms into one framework; therefore, we can capture more useful information in the temporal pooling layer. Experiments are carried out using the NIST SRE16 and SRE18 evaluation datasets. The results demonstrate the effectiveness of the GCNN and show that the proposed gated-attention statistics pooling can further improve the performance.
\end{abstract}

\section{Introduction}

Speaker verification (SV) is a task to verify a person's claimed identity from speech signals. During the last decade, the i-vector \cite{dehak2011front} combined with the probabilistic linear discriminant analysis (PLDA) framework \cite{kenny2010bayesian} has become the dominant approach in this field. The i-vector algorithm serves as the front-end, which converts a variable-length utterance to a low dimensional vector, and the PLDA algorithm is an effective backend classifier.

With the great success of deep neural networks (DNNs) in machine learning fields, more attention has been drawn to the use of DNNs to extract i-vector similar vectors, known as speaker embeddings. Many novel DNN embedding-based systems have been proposed, and they have achieved comparable or even better performance compared with the traditional i-vector paradigm \cite{variani2014deep, heigold2016end, snyder2017deep, bhattacharya2017deep, zhang2017end, snyder2018x, bredin2017tristounet, Gao2018}.

In most DNN embedding systems \cite{snyder2017deep, zhang2017end, snyder2018x, bredin2017tristounet, Gao2018}, an input utterance with a variable length is first fed into several frame-level layers to obtain high-level feature representations. The frame-level layers are usually modeled by recurrent neural networks (RNNs) \cite{bredin2017tristounet}, convolution neural networks (CNNs) \cite{zhang2017end, Gao2018} or time-delay neural networks (TDNNs) \cite{snyder2017deep, snyder2018x}. Next, a pooling layer maps all frames of the input utterance into a fixed-dimensionality vector, and the speaker embedding is generated from the following stacked fully connected layers. Furthermore, the structure optimization of the networks can help to extract better speaker-discriminant embedding. Snyder et al. employed statistics pooling to replace the commonly used average pooling in x-vector systems \cite{snyder2017deep, snyder2018x} in which the mean and standard deviation are connected together to form the utterance-level representation. The attentive statistic pooling \cite{Zhu2018, Okabe2018} is further proposed to improve the performance of the x-vector system.

Recently, the gating mechanism has proven to be useful for the natural language processing (NLP) task. Inspired by the gated linear units \cite{pmlr-v70-dauphin17a} and the long-short term memory (LSTM) network \cite{hochreiter1997long}, Peixin et al. proposed a gated convolution neural network (GCNN) \cite{Chen2018} for sentence matching and demonstrated its superiority over the CNN and RNN on sentence matching tasks. The gating mechanism combined with the memory cell is introduced in the GCNN embedding layers to capture representations that contain important information. On the other hand, the gating mechanism combined with the attention mechanism also demonstrates its effectiveness for spoken-language understanding \cite{li2018self}.

In this paper, we investigate gating mechanisms for the x-vector embedding system. More specifically, the stacked GCNN embedding layers are employed as the frame-level layers for extracting more expressive feature representations. In addition, we also apply the gating mechanism in the attention pooling layer and propose a novel gated-attention statistics pooling layer in which the attentive weights of the input frames are modeled with gating mechanism for attentive statistic pooling. We evaluate our experiments on the NIST SRE16 evaluation dataset \cite{sadjadi20172016} and the Call My Net2 (CMN2) part of NIST SRE18 evaluation dataset \cite{NIST2018}. The experimental results show the application of the GCNN for text-independent speaker verification can improve the performance of the DNN embedding system. Additionally, the proposed gated-attention statistics pooling offers further improvement over the conventional attentive statistics pooling approach.

The remainder of this paper is organized as follows. Section 2 introduces our x-vector baseline and the attentive statistics pooling. Section 3 describes the application of the GCNN, as well as the proposed gated-attention statistics pooling strategy. Section 4 presents the experimental setup and the results of this study. In Section 5, we summarize our work.

\section{Baseline network architecture}
\subsection{X-vector baseline}
The network architecture of our x-vector baseline system is similar to that described in \cite{snyder2018x}. The first five TDNN (or 1-dimensional dilated CNN) layers are stacked to extract the frame-level DNN features. The TDNN layers with dilation rates of 2 and 4 are used for the second and third layers, respectively, while the others retain the dilation rate of 1. The kernel sizes of these five layers are 5, 3, 3, 1 and 1, respectively.
The final frame-level output vectors of the whole variable-length utterance are aggregated into a fixed segment-level vector through the statistics pooling layer. The mean and standard deviation are calculated and then concatenated for statistics pooling. Two additional fully connected layers followed with a softmax layer are used to predict speaker labels. Once the DNN is trained, the output of the linear affine layer on top of the statistics pooling is extracted as the speaker embedding.

\subsection{Attentive statistics pooling}

Instead of treating the output representations from the last frame-level layer equally in the statistics pooling, the attention mechanism is applied to weight the more speaker-discriminative frames in the input utterance \cite{Zhu2018, Okabe2018}.

A single-head attention \cite{Zhu2018} strategy is applied in the x-vector baseline system. Suppose ${\bf{h}}_t$ is the hidden representation of the $t{\rm{ - th}}$ input frame below the attention layer. The attention weight ${\alpha _t}$ for ${\bf{h}}_t$ can be calculated as
\begin{equation}
\begin{array}{l}
\label{eq1}
{e_t} = {\bf{w}}_2^Tf({{\bf{W}}_1}{{\bf{h}}_t})\\
{\alpha _t} = \frac{{\exp ({e_t})}}{{\sum\limits_\tau  {\exp ({e_\tau })} }}
\end{array}
\end{equation}
where the ${{\bf{W}}_1}$ and ${{\bf{w}}_2}$ are the learned parameters, and $f$ is the ReLU function. Then, the weighted mean vector ${\bf{u}}$ and the standard deviation vector ${\bm{\delta }}$ are computed as follows
\begin{equation}
\begin{array}{l}
{\bf{u}} = \sum\limits_t {{\alpha _t}{{\bf{h}}_t}} \\
{\bm{\delta }} = \sqrt {\sum\limits_t {{\alpha _t}{{\bf{h}}_t} \odot {{\bf{h}}_t}}  - {\bf{u}} \odot {\bf{u}}}
\end{array}
\end{equation}
where $\odot$ is the elementwise multiplication. Finally, we concatenate the ${\bf{u}}$ and ${\bm{\delta }}$ to perform the attentive statistics pooling.
\section{DNN with gating mechanisms}

In this section, we will employ the gating mechanisms for the x-vector system in two ways. As depicted in Figure 1, we first use four GCNN \cite{Chen2018} layers to replace the first four TDNN layers for extracting the frame-level representations. On the other hand, we propose the gated-attention statistics pooling as an alternative attention method for aggregating the frame-level vectors. The remaining part is similar to that of the x-vector baseline.

\subsection{Frame-level GCNN layers}

The GLU \cite{pmlr-v70-dauphin17a} uses the output gate to control what information should be propagated through the convolutional layers. To better control the path through which information flows in the hierarchical structure, we adopt both output and forget gates in the GCNN. In more detail, for the $L{\rm{-th}}$ frame-level layer, the GCNN can be described as follows:
\begin{equation}
\label{eq3}
\begin{array}{l}
{\bf{o}}_t^L = \sigma ({\bf{h}}_t^{L - 1}(c)*{{\bf{W}}_o} + {{\bf{b}}_o})\\
{\bf{f}}_t^L = \sigma ({\bf{h}}_t^{L - 1}(c)*{{\bf{W}}_f} + {{\bf{b}}_f})\\
{\bf{g}}_t^L = \tanh ({\bf{h}}_t^{L - 1}(c)*{{\bf{W}}_g} + {{\bf{b}}_g})\\
{\bf{c}}_t^L = {\bf{f}}_t^L \odot {\bf{c}}_t^{L - 1} + (1 - {\bf{f}}_t^L) \odot {\bf{h}}_t^{L - 1}\\
{\bf{h}}_t^L = {\bf{o}}_t^L \odot {\bf{g}}_t^L + {\bf{c}}_t^L
\end{array}
\end{equation}
where $*$ is the convolution operation and $\sigma$ is the sigmoid function, ${{\bf{W}}_o}$, ${{\bf{W}}_f}$ and ${{\bf{W}}_g}$ are convolution parameters, while ${{\bf{b}}_o}$, ${{\bf{b}}_f}$ and ${{\bf{b}}_g}$ are bias, and ${\bf{o}}_t^L$, ${\bf{f}}_t^L$ and ${\bf{c}}_t^L$ are the output gate, forget gate and memory cell for the $t{\rm{-th}}$ frame in layer $L$, respectively. If the dimensions of ${\bf{c}}_t^L$ and the candidate ${\bf{g}}_t^L$ are not matched in Eq. \ref{eq3}, we can perform a linear projection. ${\bf{h}}_t^{L - 1}(c)$ is the input, which is formed by concatenating ${\bf{h}}_t^{L - 1}$ with its nearby frames in the previous layer or input acoustic features. Here, we apply the GCNN in a dilated way like the TDNN in the baseline system. As shown in Figure 2, when the kernel width is 3 and the dilation rate is 2, ${\bf{h}}_t^{L - 1}(c)$ can be written as
\begin{equation}
\label{eq4}
{\bf{h}}_t^{L - 1}(c) = \{ {\bf{h}}_{t - 2}^{L - 1},{\bf{h}}_t^{L - 1},{\bf{h}}_{t + 2}^{L - 1}\}
\end{equation}

The GCNN optionally allows information flow through the hierarchical structure using the gating mechanism. The output gate ${\bf{o}}_t^L$ modulates the output information of the candidate convolution ${\bf{g}}_t^L$, which contains a larger temporal context. On the other hand, the memory cell ${\bf{c}}_t^L$ can be treated as a modified residual learning that contains a smaller temporal context modulated by the forget gate ${\bf{f}}_t^L$. Therefore, the frame-level representations will benefit from both larger and smaller temporal contexts, which are further regulated through the gating mechanism.
\begin{figure}[t]
  \centering
  \includegraphics[width=0.6\linewidth,height=0.6\linewidth]{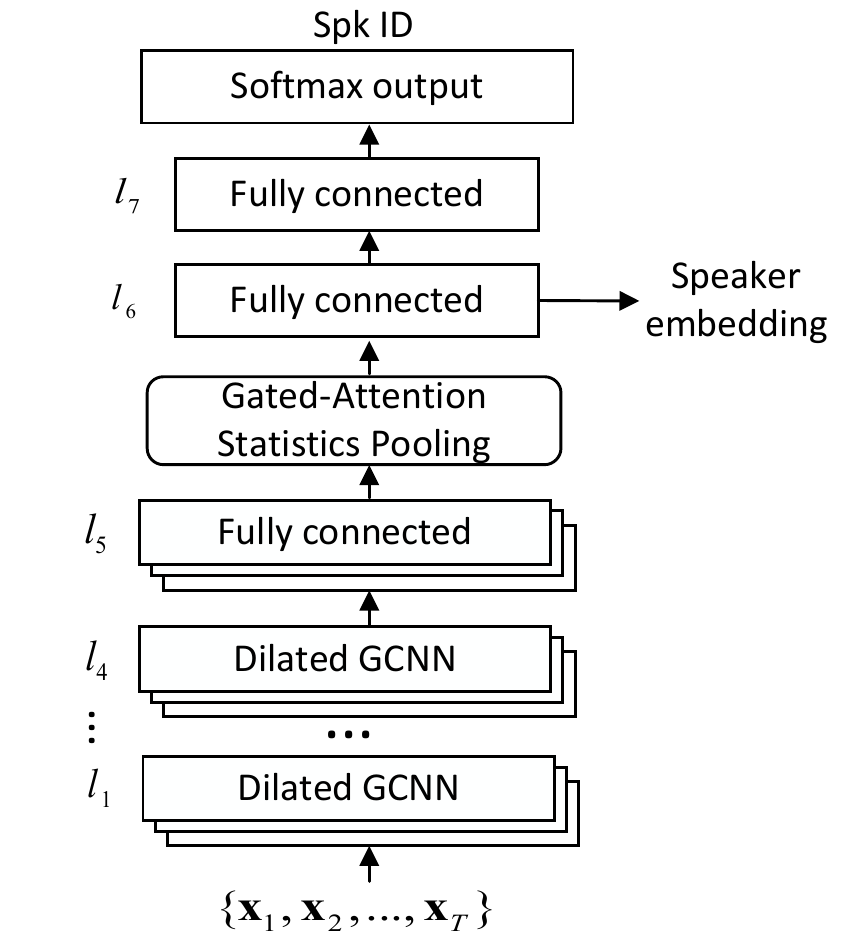}
  \caption{Network architecture of our proposed DNN.}
  \label{fig:our proposed embedding DNN}
\end{figure}
\begin{figure}[t]
  \centering
  \includegraphics[width=0.65\linewidth,height=0.45\linewidth]{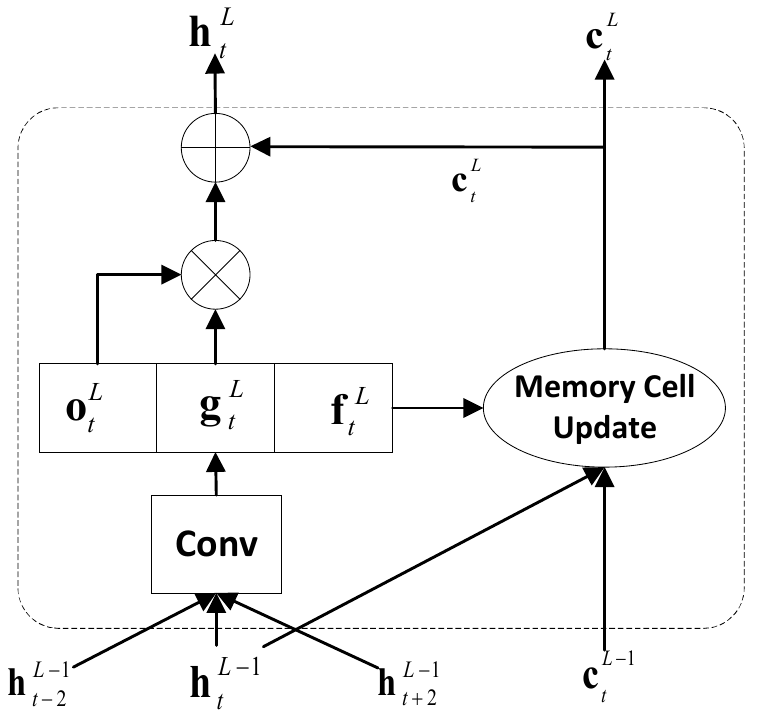}
  \caption{Structure of the GCNN unit.}
  \label{fig:Structure of the GCNN unit}
\end{figure}
\subsection{Gated-attention statistics pooling}

We propose a novel gated-attention statistics pooling in which the gating mechanism is introduced into the attention pooling layer to enhance the ability to extract speaker-discriminative information from the last frame-level layer. The gated-attention statistics pooling is depicted in Figure 3. Suppose that the last frame-level layer is the $S{\rm{-th}}$ layer. An output gate is calculated to control the output of the $S{\rm{-th}}$ hidden layer.
\begin{equation}
\label{eq5}
\begin{array}{l}
\widetilde {\bf{e}}_t^S = {\bf{h}}_t^{S - 1}(c)*{{\bf{W}}_s} + {{\bf{b}}_s}\\
{\bf{o}}_t^S = \sigma (\widetilde {\bf{e}}_t^S)
\end{array}
\end{equation}
where ${{\bf{W}}_s}$ and ${{\bf{b}}_s}$ are the weight and bias parameters. The $t{\rm{ - th}}$ output vector of the $S{\rm{-th}}$ layer is as follows.
\begin{equation}
{{\bf{z}}_t} = {\bf{o}}_t^S \odot {\bf{h}}_t^S
\end{equation}
Such a gating mechanism regulates each element of the $t{\rm{ - th}}$ output ${{\bf{z}}_t}$ with the corresponding element of the output gate ${\bf{o}}_t^S$. In other words, each element of the output gate, which is normalized between 0 and 1 with the sigmoid function, can be viewed as a kind of local attention to the corresponding ${{\bf{z}}_t}$. The mean of all the elements in $\widetilde {\bf{e}}_t^S$ can measure the importance of the $t{\rm{ - th}}$ frame-level representation. With the softmax function, the attention weight for ${{\bf{z}}_t}$ can be calculated as
\begin{equation}
{\widetilde \alpha _t} = \frac{{\exp ({\rm{mean}}(\widetilde {\bf{e}}_t^S))}}{{\sum\limits_\tau  {\exp ({\rm{mean}}(\widetilde {\bf{e}}_\tau ^S))} }}
\end{equation}
Finally, the weighted mean and the weighted standard deviation of the gated-attention pooling layer are computed by
\begin{equation}
\begin{array}{l}
\widetilde {\bf{u}} = \sum\limits_t {{{\widetilde \alpha }_t}{{\bf{z}}_t}} \\
\widetilde {\bm{\delta }} = \sqrt {\sum\limits_t {{{\widetilde \alpha }_t}{{\bf{z}}_t} \odot {{\bf{z}}_t}}  - \widetilde {\bf{u}} \odot \widetilde {\bf{u}}}
\end{array}
\end{equation}

In the gated-attention statistics pooling, from another viewpoint, the attention score for ${{\bf{z}}_t}$ is further divided into the output gate for alleviating the unimportant information among its elements. Therefore, the gated-attention statistics pooling provides a unified framework for the attention and gating mechanisms in which the former works at the frame level, and the latter can be considered as an elementwise attention. The gated-attention statistics pooling enhances both the attention and gating mechanisms by using the shared weights and helps to extract more speaker-discriminative embeddings.

\section{Experiments and analysis of the results}
\begin{figure}[t]
  \centering
  \includegraphics[width=0.7\linewidth,height=0.55\linewidth]{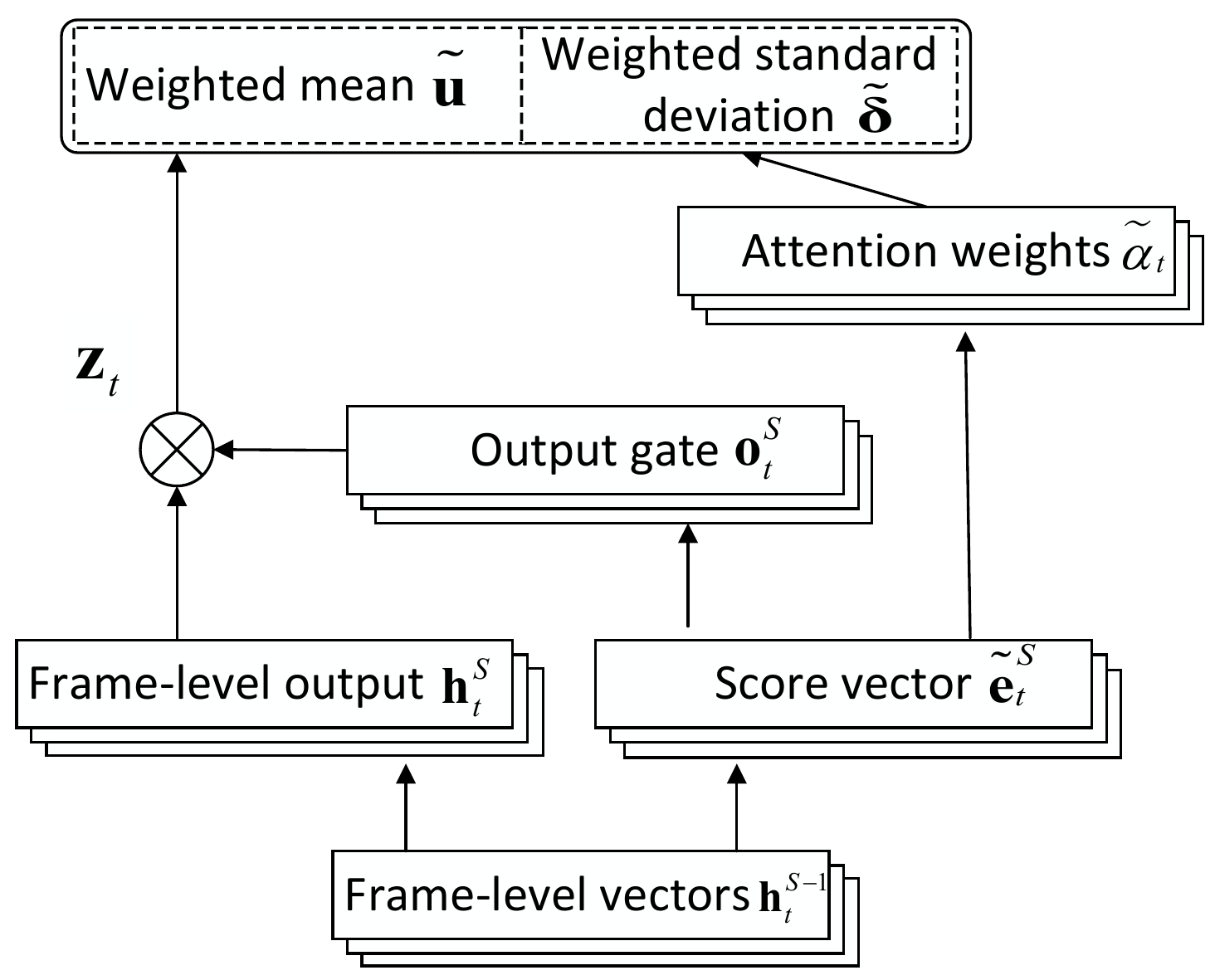}
  \caption{Gated-attention statistics pooling.}
  \label{fig:gated-attention statistics pooling}
\end{figure}
\subsection{Experimental settings}
\renewcommand\arraystretch{0.825}
\begin{table*}[th]
  \caption{Comparison results of the different systems without data augmentation}
  \label{tab:table1}
  \centering
  \setlength{\tabcolsep}{3mm}
  \begin{tabular}{cccccccccc}
    \specialrule{0.1em}{1pt}{1pt}
    & \multicolumn{2}{c}{SRE16, Pooled} & \multicolumn{2}{c}{SRE16, Cantonese} & \multicolumn{2}{c}{SRE16, Tagalog} & \multicolumn{2}{c}{SRE18, CMN2}\\
    \cmidrule(r){2-3} \cmidrule(r){4-5} \cmidrule(r){6-7} \cmidrule(r){8-9}
    system      & EER\%& minDCF & EER\%& minDCF & EER\%& minDCF & EER\%& minDCF \\
    \specialrule{0.1em}{1pt}{1pt}
    X\_TDNN             &        8.04&	0.635&	4.66&	0.429&	11.41&	0.802&	9.91&	0.622&\\
    X\_GCNN             &        7.77&	0.598&	4.40&	0.401&	11.19&	0.762&	9.35&	0.613&\\\hline
    X\_TDNN+Att         &        8.00&	0.628&	4.60&	0.423&	11.39&	0.790&	9.47&	0.632&\\
    X\_GCNN+Att         &        7.68&	0.596&	4.28&	0.397&	11.07&	0.757&	9.43&	0.609&\\
    X\_GCNN+GAtt        &        7.48&	0.585&	4.24&	0.397&	10.70&	0.742&	9.12&	0.596&\\
    X\_GCNN\_Fusion     &        \bf{7.04}&	\bf{0.568}&	\bf{3.89}&	\bf{0.374}&	\bf{10.19}&	\bf{0.734}&	\bf{8.69}&	\bf{0.581}\\

    \specialrule{0.1em}{1pt}{1pt}
  \end{tabular}
\end{table*}

\begin{table*}[th]
  \caption{Comparison results of the different systems with data augmentation}
  \label{tab:table2}
  \centering
  \setlength{\tabcolsep}{3mm}
  \begin{tabular}{cccccccccc}
    \specialrule{0.1em}{1pt}{1pt}
    & \multicolumn{2}{c}{SRE16, Pooled} & \multicolumn{2}{c}{SRE16, Cantonese} & \multicolumn{2}{c}{SRE16, Tagalog} & \multicolumn{2}{c}{SRE18, CMN2}\\
    \cmidrule(r){2-3} \cmidrule(r){4-5} \cmidrule(r){6-7} \cmidrule(r){8-9}
    system      & EER\%& minDCF & EER\%& minDCF & EER\%& minDCF & EER\%& minDCF \\
    \specialrule{0.1em}{1pt}{1pt}
    X\_TDNN             &        7.63&	0.597&	4.20&	0.412&	11.02&	0.756&	8.67&	0.579&\\
    X\_GCNN             &        7.35&	0.565&	4.23&	0.387&	10.55&	0.733&	8.35&	0.566&\\\hline
    X\_TDNN+Att         &        7.47&	0.596&	4.18&	0.422&	10.76&	0.757&	8.67&	0.573&\\
    X\_GCNN+Att         &        7.19&	0.562&	3.86&	0.385&	10.53&	0.725&	8.20&	0.556&\\
    X\_GCNN+GAtt        &        7.24&	0.560&	4.07&	0.388&	10.44&	0.721&	8.14&	0.550&\\
    X\_GCNN\_Fusion     &        \bf{6.82}&	\bf{0.545}&	\bf{3.67}&	\bf{0.371}&	\bf{9.99}&	\bf{0.710}&	\bf{7.83}&	\bf{0.537}\\

    \specialrule{0.1em}{1pt}{1pt}

  \end{tabular}
\end{table*}

\subsubsection{Training data and evaluation metric}
Our experiments are carried out in both the NIST SRE16 and SRE18 evaluation datasets. For the NIST SRE18 dataset, we only consider the CMN2 portion of the evaluation dataset. The training data mainly consists of the telephone speech (with a small amount of the microphone speech) from the NIST SRE2004-2010, Mixer 6 and Switchboard datasets. We also use the data augmentation techniques described in \cite{snyder2018x}, which employ the babble, music, noise and reverb augmented data to increase the amount and diversity of the existing training data. In summary, there are a total of 205,600 recordings, including approximately 120,000 randomly selected augmented recordings. Note that the training data for the NIST SRE16 evaluation is consistent with that for the NIST SRE18.

The performance is evaluated in terms of equal error rate (EER) and the minimal detection cost function (minDCF) calculated using the SRE16 and 18 official scoring software.

\subsubsection{Input features}

We select the 39-dimensional perceptual linear predictive (PLP) features containing delta and delta-delta coefficients as the input acoustic features. Each PLP feature is extracted from the speech signal of a 25 ms window with 10 ms frame shift. We employ the voice activity detection (VAD) to select the valid speech frames from the utterances, and the PLP features are cepstral mean and variance normalized along the whole input utterance before being fed into the embedding network.

\subsubsection{Model configuration}
We mainly compare six systems as stated as follows.

{\textbf{X\_TDNN}}: This is the x-vector baseline system without an attention mechanism. The number of hidden nodes for the first four frame-level layers is 512, while that number is 1500 for the last frame-level layer. Each of the two fully connected layers ${l_6}$ and ${l_7}$ has 512 nodes. All the nonlinear activation functions of hidden layers are ReLU. Each of the layers from ${l_1}$ to ${l_7}$ is followed by batch normalization \cite{pmlr-v37-ioffe15}. The dropouts \cite{srivastava2014dropout} and L2 weight decay are also used to prevent overfitting.

\textbf{X\_GCNN}: The GCNN layers are employed in the x-vector system as described above. The kernel size and dilation rate of each layer in the GCNN-based system are the same as those in the baseline system. The hidden size of the first four layers is 256 instead of 512 for computational efficiency. The remaining network structure is the same as that of the X\_TDNN baseline.

\textbf{X\_TDNN+Att}: The abovementioned attentive statistics pooling in Section 2.2 is used instead of the original statistics pooling in the x-vector baseline \cite{Zhu2018}, and this method achieves better performance than the original method.

\textbf{X\_GCNN+Att}/\textbf{GAtt}: For these two systems, the attentive statistics pooling and proposed gated-attention statistics pooling are employed in the GCNN-based system, respectively. The dimensionality of ${{\bf{w}}_2}$ in Eq. \ref{eq1} is set to 256 so that the number of parameters in the attentive statistics pooling is comparable to that in the gated-attention statistics pooling.

\textbf{X\_GCNN\_Fusion}: The complementarity between the above two different attention pooling methods is also investigated here. We only report the results using the score fusion of the X\_GCNN+Att and X\_GCNN+GAtt with equal weights.

\subsubsection{Network training}

The deep embedding networks are implemented with the TensorFlow toolkit \cite{abaditensorflow}. The network is trained on approximately 10-second long chunks. We use the Adam optimizer \cite{kingma2014adam} with an initial learning rate of 0.00015 and decay the learning rate based on the validation set.

\subsubsection{PLDA backend}
We use the Kaldi toolkit to implement the PLDA backend similar to that in \cite{zeinali2018improve}. For both NIST SRE16 and SRE18, the DNN embeddings are centered using the unlabeled development data and projected using LDA. The PLDA model is trained and then adapted to the unlabeled data through the unsupervised adaptation in Kaldi.

\subsection{Results and analysis}
Since it is meaningful to observe the performance of both systems with and without data augmentation, we will report the performance of both systems.
\subsubsection{Results}
Table 1 presents the performance of the different DNN embedding systems without data augmentation in the NIST SRE16 and SRE18. It can be observed that our GCNN-based systems outperform the TDNN baseline in all evaluation conditions in terms of both EER and minDCF, regardless of whether the systems are equipped with an attention mechanism or not. Clearly, it demonstrates the effectiveness of our frame-level GCNN layers. Moreover, X\_GCNN+GAtt achieves a better performance compared with X\_GCNN+Att and provides an 8\% relative improvement over X\_TDNN+Att on Cantonese in terms of EER. This observation shows our proposed gated-attention statistics pooling can further enhance the performance of GCNN-based systems over the original attentive statistics pooling. Among all the above systems, the fused system achieves the best performance and improves on the attention-based TDNN baseline by 15\% on Cantonese and 11\% on Tagalog in NIST SRE16 in terms of EER.

The performance of the systems with data augmentation is reported in Table 2, from which we can draw a similar conclusion as from Table 1. In this case, although X\_GCNN+GAtt is comparable with X\_GCNN+Att in NIST SRE16, it is still slightly better than X\_GCNN+Att in NIST SRE18 in EER and minDCF. Again, the GCNN-based fusion system gives the largest improvement over other systems. In terms of both EER and minDCF, it is 12\% better than X\_TDNN+Att on Cantonese in NIST SRE16. For NIST SRE18, a 10\% relative improvement is achieved compared with the attention-based baseline system in terms of EER. These results make clear that the TDNN-based x-vector system is enhanced significantly with our introduced gating mechanisms.
\subsubsection{Further analysis}
Here, we investigate the effect of removing different parts in the gated-attention statistics pooling. Table 3 lists the pooled results in SRE16 and the CMN2 results in SRE18 of the systems without augmentation. We set up X\_GCNN+Gonly in which the attention mechanism is removed such that only the output gate works in the pooling layer. We observe a certain decrease in the performance compared with X\_GCNN+GAtt. We can also observe the performance degradation in the X\_GCNN+Aonly system in which the output gate is removed. This observation indicates that both the attention and gating mechanisms contribute to the final performance in the gated-attention statistics pooling layer.

\begin{table}[th]
  \caption{Performance comparisons of removing different parts in gated-attention statistics pooling}
  \label{tab:table3}
  \centering
  \setlength{\tabcolsep}{0.5mm}
  \begin{tabular}{cccccc}
    \specialrule{0.1em}{1pt}{1pt}
    & \multicolumn{2}{c}{SRE16, Pooled}  & \multicolumn{2}{c}{SRE18, CMN2}\\
    \cmidrule(r){2-3} \cmidrule(r){4-5}
    system      & EER\%& minDCF & EER\%& minDCF  \\
    \specialrule{0.1em}{1pt}{1pt}
    X\_GCNN+Gonly           &        7.85&	0.588&	9.27&	0.615&\\
    X\_GCNN+Aonly           &        7.75&	0.594&	9.28&	0.602&\\
    X\_GCNN+GAtt            &        \bf{7.48}&	\bf{0.585}&	\bf{9.12}&	\bf{0.596}&\\

    \specialrule{0.1em}{1pt}{1pt}

  \end{tabular}
\end{table}

\section{Conclusions}
In this study, we employ gating mechanisms in the DNN embedding system. More specially, the GCNN is introduced into the frame-level layers where the output representations of the GCNN layer are carefully modulated through both the output gate and memory cell. Such a mechanism helps to obtain more expressive features. Furthermore, we propose a gated-attention statistics pooling in which the attention model is enhanced by the gating mechanism by sharing the same weights with the output gate. The experimental results demonstrate that GCNN-based systems outperform the TDNN-based x-vector baseline, and our proposed gated-attention statistics pooling provides further improvement over the original attentive statistics pooling in GCNN-based embedding systems. Moreover, the proposed pooling method has obvious complementarity with the attentive statistics pooling, and the fused system achieves the best performance among all the mentioned systems.
\section{Acknowledgements}
This work was partially funded by the National Natural Science Foundation of China (Grant No. U1836219) and the National Key Research and Development Program of China (Grant No. 2016YFB100 1303).

\bibliographystyle{IEEEtran}

\bibliography{mybib}


\end{document}